\documentclass{chjaa}
\usepackage{graphicx}
\renewcommand\deg{\ifmmode^\circ\else$^\circ$\fi}
\begin{document}


\title{A Possible Periodicity in the Radio Lightcurves of 3C454.3}

\author{S.J.~Qian\inst{1,2} \and N.A.~Kudryavtseva\inst{1,4,5}
        \and S.~Britzen\inst{1}
        \and T.P.~Krichbaum\inst{1}
       \and A.~Witzel\inst{1}  \and J.A.~Zensus\inst{1}
       \and M.F.~Aller\inst{3} \and H.D.~Aller\inst{3}
       \and X.Z.~Zhang\inst{2}}

\offprints{S.J.~Qian, email: rqsj@bao.ac.cn}

\institute{Max-Planck-Institut f\"ur Radioastronomie, Auf dem H\"ugel 69,
        53121 Bonn, Germany
    \and    National Astronomical Observatories, Chinese Academy of Sciences,
        Beijing 100012, China
   \and Astronomy Department, University of Michigan, Ann Arbor, MI 48109, USA
  \and Astronomical Institute of St.-Petersburg State University,
     Petrodvoretz, St.-Petersburg, Russia
  \and International Max-Planck Research School for Radio and Infrared
  Astronomy at the Universities of Bonn and Cologne}

\date{2006.12.5}

\abstract{During the period 1966.5--2006.2 the 15GHz and 8GHz lightcurves
 of 3C454.3 (z=0.859) show a quasi-periodicity of $\sim$12.8 yr
 ($\sim$6.9 yr in the rest frame of the source) with a double-bump structure.
   This periodic behaviour
  is interpreted in terms of
 a rotating double-jet model in which the two jets are created from
 the black holes in a binary system and  rotate with the period
 of the orbital motion.
 The periodic variations in the radio fluxes
 of 3C454.3 are suggested to be mainly due
 to the lighthouse effects (or the
  variation in Doppler boosting) of the precessing jets which are caused
  by the orbital motion. In addition, variations
 in the mass-flow rates accreting onto
   the black holes may be also involved.\\
 \keywords{radio continuum: galaxies-Quasars: individual: 3C454.3}
} \maketitle
 \section{Introduction}
  In recent years many works have been devoted
 to the study of the periodic
  properties discovered in extragalactic sources,
 especially in Blazars. These
  include the periodic (or quasi-periodic) variations of flux density (or
 luminosity) in optical bands (for example, for OJ287:
 Sillanp{\"a}{\"a} et al. \cite{Si88}, Lehto \& Valtonen \cite{Le96},
    Valtonen et al. \cite{Va99},
 Villata et al. \cite{Vi98}, Valtaoja et al. \cite{Va00},
 Valtonen \cite{Va06a}, \cite{Va06b}; AO\,0235+16: Raiteri et al. \cite{Ra01});
  periodic variations of flux density in radio
  bands (for example, for AO0235+16: Raiteri et al. \cite{Ra01}, 3C454.3:
 Ciaramella et al. \cite{Ci04}, Kudryavtseva \& Pyatunina \cite{Ku06});
  the periodic changes of
  the position angle of the ejection
  of superluminal knots (for example, Qian et al. \cite{Qi01},
   Britzen et al. \cite{Br01},
 Abraham \& Carrara \cite{Ab00}, Lobanov \& Roland \cite{Lo05});
 periodic changes
  of the trajectory of superluminal
  knots (for example, for 3C345: Qian et al. \cite{Qi91},
  Lobanov \& Roland \cite{Lo05},
  Klare et al. \cite{Kl05}), etc.
 Up to now, the most reliable periodic phenomenon observed in blazars
 may be the 12 year
 quasi-periodic optical
 variations
 of the BL Lac object OJ287 which has been convincingly
 determined in the more than 100 year long
 records of optical observations since 1890.\\
 In the interpretations of these periodic phenomena various
 binary balck hole models
 have been invoked. For the periodic optical variations in OJ287  all the
 proposed models invoke a binary system with two supermassive
 black holes. But the mechanisms for the production of the optical
 outbursts can be divided into two classes: accretion models and lighthouse
 models.
 \begin{itemize}
 \item Accretion models\\
 These models (Lehto \& Valtonen \cite{Le96}, Valtonen \cite{Va06a},
   \cite{Va06b})
 propose that the optical
 periodicity is caused by a precessing binary black hole system of
  two supermassive black holes with a
 large eccentric orbit.
 The observed periodic 'superflares'  with double peaks are suggested to be
 due to the
 crossings of the secondary black hole into the accretion disk
 of the primary black hole during the pericenter passage.
 Thus in this class of models the optical emission is thermal (bremsstrahlung)
 and the
 observed flux increase
 reflects the enhanced accretion related to the disk crossings and  the
 general enhancement of the accretion rate during the pericenter passage.
 Doppler boosting is not taken into account. Detailed models for a
 binary black hole system have been
 proposed to fit the past  optical records for OJ287 and predict the future
 optical events. If the timing of the optical activity (superflares) in
 OJ287 can be accurately predicted, then it is possible to calculate or
 determine the parameters of the binary black hole system (such as the masses
 of the black holes, the orbital parameters etc.).

 \item Lighthouse models\\
 The periodic optical flares observed in OJ287 are suggested to be caused by
 the change of the orientation of the realtivistically moving emission
 regions with respect to the line of sight, resulting in an increase of the
 Doppler boosting factor (Camenzind \& Krockenberg \cite{Ca92},
  Katz \cite{Ka97}, Villata et al.
 et al. \cite{Vi98}, Qian et al. \cite{Qi01}) and
  the apparent optical flux density
 (${\propto}{{\delta}^3}$). Katz (\cite{Ka97}) proposed that
 in a bianry black hole system
 the companion exerts a torque on the accretion disk of the primary black hole
 and results in the precessing of the relativistic jet from the primary
  sweeping across the line of sight and the periodic variation of the Doppler
 factor and thus the apparent flux density.
   Villata et al. (\cite{Vi98}) suggests
 that the two black holes of the binary system both create relativistic jets
 which are bent significantly in different directions.
 In the course of the binary's orbit motion, the directions of the
 bent parts of the jets from the two black holes rotate with the orbital
 period, resulting in periodic double-peak flares. In this class of models
 only the variation of the Doppler boosting factor plays the role, not
 considering any change in accretion in the central disk-hole systems.
 \item Alternative models\\
   Valtaoja et al. (\cite{Va00}) have argued against both
 the accretion models and
 lighthouse models for OJ287. They found that in OJ287
 the first optical flare of the double-structure is thermal (non-polarized)
 and lacks
 radio counterpart, but the second one is (polarized)
 synchrotron emission, having
 a radio counterpart. They proposed an alternative model in which the first
 optical flare is caused by the disk-crossing during the pericenter passage,
 having a regular period.
  And the second one occurs about one year later in the relativistic
 jet with an
 association with a radio outburst. For
 checking this model optical polarization measurements and VLBI studies
 of the optical-radio association are significant.
 \end{itemize}
 In this paper we will discuss the possible existence of a $\sim$13
 year periodicity in the radio lightcurve of the optically violently
 variable (OVV) quasar 3C454.3 at 8GHz and 14.5GHz in which two broad
 peaks (or bumps)
  are observed during one period. We propose that the periodicity can be
 interpreted in the frame of a binary black hole model in which two jets
 from the two black holes rotate with the period ($\sim$13 year)
 of the orbital motion. Model-fits to the lightcurves are given.
   It is shown that both the periodic
 Doppler boosting effect (lighthouse effect) and the variability of the
 accretion activity (i.e. the mass-energy inflow into the jets)
  should be taken into account in order to fully
 explain the lightcurves. We also discuss some alternative models
 and future observations for further studying the periodic phenomenon
 in 3C454.3.

 \section{Optical and radio properties of 3C454.3}
  3C454.3 was early defined as typical optically violently variable (OVV)
 quasar sometimes showing flare-ups of 2-3 magnitudes.
 Historical optical data go back to 1900 (Angione \cite{An68}).
 It has a flat spectrum in the cm-mm bands and belongs to the  category of
 blazars, showing
   strong flux variabilty over the whole electromagnetic spectrum from
  radio to $\gamma$-ray (Villata et al. \cite{Vi06},
  Pian et al. \cite{Pi06}).\\
 \subsection{Optical variability}
 As shown by Villata et al. (\cite{Vi06}), the optical variability of 3C454.3
 (cf. their Fig.1 (upper panel)) shows two different phases of activity.
 Since 1966 till 2001 (during a 35 year period) the observed
   optical variability
  is moderate
 with $m_R$ in the range of $\sim$15--17 magnitude. Before 1966  few data
 are present. The historical data during the 60 year
 period (from 1900
 to 1960) given by Angione (1968) show a peak magnitude of $m_B$$\sim$15
 (equivalent to $m_R$$\sim$14) at 1953.6, the most rapid variability
 was about 0.4mag/day and the total variation $\sim$2 mag (17 to 15) in B-band.
  From 2001 the amplitude of the optical variations
 starts to increase and 3C454.3 was observed at $m_R$$\sim$13.5 in
  August/September 2001 (Fuhrmann et al. \cite{Fu06}). Recently, It was
 observed at $m_R$$\sim$12 on 2005.35 (May 9.36),
  becoming the brightest quasar in the sky (Villata et al.\cite{Vi06}).
 Thanks to the WEBT group, a multi-
 frequency campaign of observations was conducted in radio and optical bands
 during this high activity
 (see Villata et al. \cite{Vi06}; Pian et al. \cite{Pi06},;
 Fuhrmann et al. \cite{Fu06}).\\
  With regard to the study of the radio periodicity in 3C454.3
  in this paper (see below), the unprecedented optical outburst in 2005 and the
 associated huge millimeter outburst are very significant, because they
  show us that a quasi-periodicity of $\sim$12-13 year is really existing.
 (Here we  would mention: by visual inspection of
 the radio lightcurves
 at 8GHz and 15GHz during
 the period 1966--2004\footnote{This research has made use of data from the
 University of Michigan Radio Astronomy Observatory, which has been supported
 by the University of Michigan and the National Science Fundation.}
  and without accessing to any other information, we
  predicted the next peak of its activity to be
 occurring in 2006--2007.  The
  optical outburst at 2005.35 which we later learned  from literature
 just verified our prediction, considering that radio outbursts
   usually delay to optical ones by about $\sim$1 year).\\
  Quasi-periodicities in optical variations of 3C454.3 have been studied.
 Webb et al. (\cite{We88})  analysed the B-light curves during the
  period (1971--1985) and obtained three (weak) periods: 6.4, 3.0
  and 0.8 years.
 Su (\cite{SuC01}) used an optical dataset (B-magnitude) including the years
 from 1900 to 1996 and found a period of 12.4 yr. These previous results
 for the optical variations (the two periods of 6.4 and 12.4 yr)
 are unexpected to be consistent with the periods
 obtained for the radio variations (see below).\\
  The optical emission from 3C454.3 is polarized during
 outburst and nonoutburst
 periods and also when it reaches minimum brightness. It
 is a high polarization quasar, which
 was observed to be in high optical polarization states
 with a maximum polarization degree of $\sim$16$\%$.
 Its high polarization states associate
 with strong variability, not with high brightness
   (Visvanathan et al. \cite{Vi73},
 Angel \& Stockman \cite{Ang80}). Sometimes it is in very
 low polarization states
 of p$\sim$0$\%$
 (Moore \& Stockman \cite{Mo81}, Pollock et al. \cite{Pol79}). These
 poalrization observations seem to show that the optical variations in both
 high and low states  originate from the synchrotron emission of the jets.
 This is different from OJ287 (Valtaoja et al. \cite{Va00})
 and we will propose a
 double jet model to fit the radio ligh curves (see below).
 \subsection{Radio variability and quasi-periodicity}
 3C454.3 is a radio quasar with a flat spectrum. As shown  in Fig.1,
 its radio lightcurves at 8GHz and 14.5GHz over nearly 40 years
 show several outbursts. As well known,
 the existence of periodicity
 in radio light curves is still a controversial issue, even in the case
 for OJ287 in which periodic optical variations have been discovered
 (Sillanp{\"a}{\"a} et al. \cite{Si88},
  Valotonen \cite{Va06a}, \cite{Va06b}). However,
   Ciaramella et al. (\cite{Ci04})
 have made detailed analysis of the radio data (1970--1999)
  for 3C454.3 at five
 frequencies (4.8, 8.0, 14.5, 22 and 37GHz) by using
 autocorrelation method and pointed out that a strong and
  consistent periodicity of about six
 years ($\sim$6.0--6.5yr)
 was found in the overall flaring behaviour. 3C454.3 was regarded as
 the best case for evidence of a radio periodicity
   in a subsample of 77 extragalactic sources
 selected from the sample of Ter{\"a}srants et al. (\cite{Te98}).\\
  By visual inspection
 of the radio lightcurves given by Ciaramella et al. (\cite{Ci04}),
  we find that,
 if we take into account the apparent difference between the amplitudes of
 the peaks,
  the quasi-periodicity may be
 $\sim$12-13 years with a second peak always weaker than the main peak.
 The unprecedented optical (and millimeter) outburst observed in 2005.35
 described above seems
 to verify this identification (the optical outburst signals the coming of
 a new main peaking period following the radio peaks at epochs 1967 (8.0GHz),
 1981 (8 and 14.5GHz)
 and 1994 (at 8 and 14.5GHz). See Figure 1.\\
 We should point out that normally in radio bands few blazars
 have been found to be periodic in their radio light curves. For example,
 even in the BL Lac object OJ287,
 its optical lightcurve has a 12 year periodic flaring with a double structure,
 but there is no periodic radio behaviour corresponding to
 the optical cycles. Especially, there are no outstanding individual radio
 outbursts corresponding to the optical double 'superflares" (Valtaoja et al.
 (\cite{Va00}). The radio outbursts consist of many short timescale
  (1-2 year) peaks overlapping on a bump which has
 a $\sim$22-25 year period (about twice of the optical period;
 see Ciaramella et al. \cite{Ci04}).
   In contrast, in 3C454.3
 the radio outbursts are outstanding, having simple profiles (especially
 for the main bumps),
 thus its
 quasi-periodicity can be easily recognized.\\
 Most recently, Kudryavtseva \& Pyatunina (\cite{Ku06})
 have studied the radio
 periodic behaviour in three sources (3C454.3, CTA102 and 3C446).
 They have made a detailed analysis
 of the radio light curves at 37, 22, 15, 8 and 4.8\,GHz for 3C454.3
 and found two periods of 6.2 and 12.4 yr. In addition,
 Li et al. (\cite{Li06})
 found aperiod of 6.2 yr. from a periodic analysis of the 37 and 22\,GHz
 light curves.
 Thus, combining the results obtained by Webb et al. (\cite{We88}),
 Su (\cite{SuC01}), Ciaramella et al. (\cite{Ci04})
  Kudryavtseva \& Pyatunina (\cite{Ku06}) and Li et al. (\cite{Li06}),
  we would come to the conclusion that
 in both radio and optical regimes there seems two quasi-periods
 ($\sim$6.2 and
 $\sim$12.4 yr) existing.\\
 For the discussion below
 we summarize the radio behaviour of 3C454.3 (shown in Figure 1) as follows.
 \begin{itemize}
 \item (1) Since 1966,
  three major radio outbursts have been observed at 14.5GHz,
 peaking at 1981.5, 1994.5 and 2006.2. Also  second
 peaks (or bumps) were observed, they are $\sim$7 years after the first
 bumps, the most noticeable one is at 1987.5, giving a double-bump
 structure during the period (1980--1991).Thus a 12--13-year periodicity
 could be identified. For the 8GHz lightcurve there are also  three
 outstanding cycles
    observed (1966.5, 1981.5, 1994), because the radio-mm outburst at 2005.6
 is not noticeable at 8GHz.
 In addition, for each cycle two broad peaks (bumps) are clearly seen and
    the second bumps are always less regular with smaller amplitudes. This
    might imply that the two bumps could originate from different regions.
 \item (2) The relation between centimeter and millimeter outbursts in 3C454.3
  is observed to be different at different epochs.
  For example, during the period of the
 optical outstanding flare in 2005, the mm-flares at 1-3mm are closely
 associated with the optical
 one and its amplitude is much larger than those at centimeter
  wavelengths (e.g., at 2.8cm; see Villata et al. \cite{Vi06},
  Krichbaum et al. in prep.). However, during the period 1990--2004,
 the mm-lightcurve is similar to the cm-lightcurve, showing a
 similar double-bump structure, but the amplitudes
 of the main mm-bumps are clearly smaller than those of the 14.5--8GHz bumps.
 Generally, the outbursts at centimeter wavelengths delay to the mm-outbursts
 , causing a difference between the periods determined at
 different frequencies.
 This complex relationship
  could be due to opacity  and evolution effects of
 the shocked regions in the curved jet.
  \end{itemize}
  \begin{figure}
  \includegraphics[width=7cm,angle=-90]{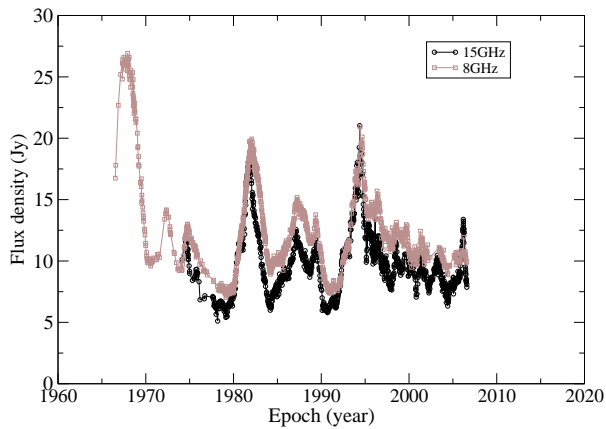}
  \caption{3C454.3: the 8 and 14.5\,GHz lightcurves,
    showing a $\sim$12-13
  year quasi-periodicity.}
  \end {figure}
  \subsection{VLBI structure}
  For proposing a model to interprete the periodic radio behaviour of 3C454.3
  VLBI observations are important. VLBI observations have been conducted
  since 1981.5 by Pauliny-Toth et al. (\cite{Pa87})
  and Pauliny-Toth (\cite{Pa98}).
  The former paper gave the 2.8cm VLBI results during the period
 1981.4--1986.2 following the 1981.5 peak, and the latter gave
 1981.6--1991.9 6cm VLBI observational results.
   These VLBI observations show that the source has a core-jet structure,
  ejecting superluminal knots following  major radio outbursts. The complex
  core was
  once observed to show a 'superluminal' increase in its size.
  The VLBI
  jet is very curved and acceleration of the motion of its knots
  was observed (apparent speed from $\sim$5c to $\sim$20c).
 However, in 3C454.3
  some features were observed to be remained stationary,
 showing their orientation
 could be directed to us (with viewing angle $\sim$$0^{\circ}$).
  Basing on the VLBI measurements, small viewing angle (less than $\sim$
 $2-5^{\circ}$) and large bulk
 Lorentz factor (larger than $\sim$20) were suggested.\\
 The results (structure and superluminal motion properties of 3C454.3)
 obtained from mm-VLBI observations (e.g. Pagels \cite{Pa04},
  Krichbaum et al. \cite{Kr96})
 are consistent with those
  obtained from cm-VLBI observations. Most recently, during the strong
 optical-mm outbutrst in 2005, 43GHz VLBI observations show variations in
 structure and changes in its flux and polarization
 (Villata et al. \cite{Vi06}).\\
 Basing on the epochs of the superluminal components ejected from the core
 of 3C454.3, Kudryavtseva \& Pyatunina (\cite{Ku06}) found that the birth
 epochs of VLBI components have similar periods as described above. The
 variation in the core  structure after the 2005 optical flare (Villata et al.
  \cite{Vi06}) seems also
 consistent with their results.
  \subsection{Radio-optical correlation}
  The relation between the radio and optical variability is complex
  for 3C454.3. In the early years, Pomphrey et al. (\cite{Po76})
  already found a
 correlation between the optical and 2.8cm radio events,
 revealing the optical
 leading the radio by $\sim$1.2 years. But later observations does not
 confirm this simple relation, i.e., generally radio outbursts were not
 observed to be associated with optical outbursts (during 1966--2004 period),
 except one or two outstanding optical flares like those
 observed in 1979 and 2005). From the observational side, the lack
  of the association is due to two reasons: (1) the range of the optical
  variability amplitude is usually small (less than $\sim$1 mag.); and (2)
 the radio outbursts are usually associated with small optical flares and
 verse versia.\\
  The general lack of correlation between the optical and radio
 behaviour suggests that the optical and radio emission originate in different
 regions and possibly misalign with respect to each other. Recently
  Villata et al.
 (\cite{Vi06})  suggest that before 2001 the radio region
   was more aligned with the line
 of sight and the radio variability was enhanced by Doppler boosting
 effects. On
 the contrary, since 2001 the optical activity increases and the 2005
 optical flare reaches a peak of $m_R$$\sim$12, an unprecedented high
 brightness. Thus in the last five years (2001-2005)
 the optical radiation region
  has a smaller
 viewing angle and enhanced by Doppler effects.
  However,
  during the period (2001--2004) the radio and mm activity  was in their
  low states. The 2005 optical flare starts  a new activity in mm-cm bands
  with the unprecedented mm-flares (peaks of $\sim$45\,Jy at 1mm and
   $\sim$26Jy at 3mm), compared with the low activity during the 1985-2001
  period (1-3mm varied in the range between $\sim$2Jy and $\sim$15Jy, when the
  observed optical activity ranging between $m_R$$\sim$15 and $\sim$16.5, also
  at  low activity). However, we should point out that in some cases
  even at low
  optical states, some close association between individual
 optical and radio
  flares can be recognized. For example, Tornikoski et al. (\cite{To94})
 found  from their optical monitoring during 1980-1993 that  the only
 prominent feature in the optical light curve was a strong flare peaking
 at $\sim$1988.7  which had its counterparts at 230, 90 and 37GHz with
 possible delays and
  might be associated with the delayed 14.5 and 8\,GHz outbursts at 14.5 and
  8\,GHz
 (Fig.1).\\
  Another feature in relation between the optical and radio flares is that the
  optical activity has considerably shorter timescales generally.
 This could be
  due to the accretion-energy supply processes (acceleration of
 electrons, radiative losses and injection rates etc.),
  not due to Doppler time
  dilution or narrower Doppler boosting profile for optical emission
 (see below).
  \section{Model-fitting}
  We have three alternatives to make model -fitting.
  \begin{itemize}
   \item (1) Binary black hole models
  as suggested by Lehto \& Valtonen (\cite{Le96}) and Valtonen (\cite{Va06a})
  for OJ287\\
   Periodicity is
  created by the pericenter passage and the disk passage of the secondary
  black hole during its orbital motion.
   This model mainly explains the double
  optical peaks in terms of the increase in the accretion into the
   primary black hole  and the crossing of the secondary black hole
  into the disk of the primary. The optical flares are considered to be
  produced near the black hole-disk system.
   The radio outbursts
  do not show a similar periodicity and are
  regarded as a phenomenon associated with the optical flaring, but produced
  in the jet (only one jet from the primary is considered). This implies that
 the intermediate processes, which dertermine the transfer of the energy from
 accretion to the energy injected into the jet, would play a significant role.
 And these processes are not clarified in these models. Therefore,
  since the periodic behaviour in 3C454.3 occurs in radio bands
 which probably is a
  phenomenon in the jet,  the binary black hole model proposed by
 Valtonen (\cite{Va06a}) can not be directly applied
  to 3C454.3. Moreover, since the optical flares in 3C454.3  seems probably
 to be synchrotron in origin,  thus how to combine the disk-passage mechanism
 and Doppler boosting effect to explain the optical flares is still
 a problem;
  \item (2) Binary black hole models with two jets which rotate due to
  orbital motion, as suggested by Villata et al. (\cite{Vi98}) for OJ287.
   Each jet produces one
   optical peak, thus forming a double optical flaring structure. The
  optical flarings are caused purely by Doppler boosting variation without
  considering the change in accretion rate and
 the mass-energy transfer into the jet.
  The optical periodicity is due to
  the periodic variation in the Doppler factor.  This model is proposed
  for explaining  optical periodicity with narrow profiles
 of optical flares ($\sim$1 year).
  In its present form the model can not be applied for 3C454.3, because in
  order to interprete the narrow width of the optical flaring profiles
  ($\sim$1 year)
  the bent parts of the two jets are required to vary in a large range
 of viewing angle
  between $\sim$$6^{\circ}$ and $\sim$8$6^{\circ}$.
  This range for the variation of
  the  viewing
  angle is too large for fitting the profiles of the radio bumps
  observed in 3C454.3. Moreover, the corresponding range for the Doppler
 beaming factor (${\delta}^3$) is extremely large (bulk Lorentz factor is taken
 to be 10, see Villata et al. \cite{Vi98}), which seems to be
 inconsistent with VLBI observations. Thus even for optical
 flares in OJ287 alternative
 mechanisms for the narrow widths (or short timescales)
 of the optical flares (e.g.
 mechanism of acceleration of electrons emitting optical synchrotron)
 should be explored.
  \item (3) Periodicity can be produced in a fully-collimated jet by helical
  motion of suplerluminal knots. This is like the  'ligthhouse' model suggested
  by Schramm et al. (\cite{Sc93}) for optical flaring in 3C345. However,
  this mechanism is only used for short-term periodic behaviour
 (timescales of several months, or at most a few years) and for a
  single peak phenomenon and thus seems not applicable to
  explain the periodicity observed in 3C454.3, where
  periodicity lasts at least 40 years.
 \end{itemize}
  Taking into consideration that in the case of 3C454.3,
  both optical and radio flares occur during  the main and secondary bump's
  period, we think that a binary black hole model with
   two jets\footnote{Two counter-jets are not involved here.}
  could be more plausible. In this case, one jet produces the main bump and
 the other jet produces the secondary bump. And
 in each curved jet there could be
   two emiting regions, one of which emits at optical and mm wavelengths and
  the other emits at cm wavelengths. The two regions could
  have different positions in the curved
   jets. The optical regions situate nearer and  the radio regions
   further from the binary system.\\
  We further assume that both the jets rotate together
  with a period (T year), which could be due to
  the orbital motion of the binary system as suggested by Villata et al.
 (\cite{Vi98}), and thus the different
  Doppler boosting profiles determine the relative
  timing of the optical flaring and
   radio flaring.
   The combination of
  the effects of accretion
  (mass-energy transfer into the jets) and  Doppler boosting
  determine the strength of the optical and radio outbursts.
  \subsection{Formulism}
   Specifically,
  we assume two jets being created from the two black holes of a binary system.
 The  original jet axes are parallel to the normal of the orbital plane (
 i.e., the orbital angular momentun). We assume that the jets precess around
 the normal due the orbital motion with the orbital period through the
 mechanism suggested by Roos et al. (\cite{Ro93}), that is,
 the vector addition of the orbital velocity of the holes and the
 velocty of he jets produce the precession of the jets in the observer's
 frame. (This assumption is different
 from that of Villata et al. (\cite{Vi98}), in which
  the outer parts of the jet are bent backwards with respect to their orbital
 motion due to  the magnetohydrodynamic interaction
 between the magnetized
 jet flows and the ambient medium).
  The bent parts which
  emit optical and radio emission
   make small angles with the normal of the orbital plane. The jet flows
 have a large Lorentz
   factor.
   We designate the bent parts of the two jets as
   component-1 (for the main radio bump) and comoponet-2 (for the secondary
    bump).
    We assume that at t=0 the observer and component-1
   are situated at the same azimuth angle (${\phi}_1$=$0^{\circ}$)
   on the orbital plane and component-2 at ${\phi}_2$=17$5^{\circ}$.
  The observer direction makes an angle $i$ with the normal
   of the orbital plane and
   component-1 and component-2 make an angle ${\psi}_1$ and ${\psi}_2$
  with the orbital normal, respectively. As described above the two
 jets rotate with
  an angular speed $\omega$=2$\pi$/$T_{obs}$ (in the observer's frame).
  We adopt a large Lorentz factor
 for the jets, as suggested by the VLBI observations (Pauliny-Toth
   et al. \cite{Pa98}).
  Then the variation with time of the viewing angles
 (${\theta}_1$ and ${\theta}_2$)
  of the two components can be written as follows.
   \begin{equation}
    {\cos{{\theta}_1}}(t)=\sin{{\psi}_1}\,\cos({\omega}t)\,\sin{i}+
                     \cos{{\psi}_1}\,\cos{i}
   \end{equation}
   \begin{equation}
    {\cos{{\theta}_2}}(t)=\sin{{\psi}_2}\,\cos({\omega}t)\,\sin{i}+
                     \cos{{\psi}_2}\,\cos{i}
   \end{equation}
    Thus we have the Doppler factors (${\delta}_1$ and ${\delta}_2$)
  and the flux densities ($S_1$ and $S_2$)
   of the components:\\
   \begin{equation}
   {{\delta}_1}(t)={[{{\Gamma}_1}(1-{{\beta}_1}{\cos{{\theta}_1}})]^{-1}}
   \end{equation}

   \begin{equation}
    {{\delta}_2}(t)={[{{\Gamma}_2}(1-{{\beta}_2}{\cos{{\theta}_2}}]^{-1}}
   \end{equation}

   \begin{equation}
    {S_1(t)}={S_{b1}}\,{{{\delta}_1}^3}
   \end{equation}

   \begin{equation}
    {S_2(t)}={S_{b2}}{{{\delta}_2}^3}
   \end{equation}
    where ${S_{b1}}$ and $S_{b2}$ are constants for normalization for
    fitting the observed light curves. We also introduce a quiescent level
   to taking account the flux density from
  the components which are not related to
   the beaming components.\\
   The parameters that we have chosen are listed in Table 1 for two models
    (taking ${\Gamma}_1$=${\Gamma}_2$): model-A and model-B.
  The only difference
  between the model-A and model-B
    is the value for ${\psi}_2$: $10^{\circ}$ for
 model-A and $5^{\circ}$
    for model-B, thus making the width of
  the Doppler profile for component-2
    to be narrower and the corresponding Doppler
 factor greater (see Figures
   2--11).
    \begin{table}
    \caption{Parameters for the two models.}
    \label{table:1}
    \centering
    \begin{tabular}{c c c}
    \hline
    Parameter  & Model-A & Model-B\\
    \hline
    $\Gamma$ & 20  & 20\\
    $T_{obs}$  & 12.8yr & 12.8yr\\
    ${\psi}_1$ & $3^{\circ}$ & $3^{\circ}$\\
    ${\psi}_2$ & 1$0^{\circ}$  & $5^{\circ}$\\
    ${\phi}_1$ & $0^{\circ}$  & $0^{\circ}$\\
    ${\phi}_2$ & 17$5^{\circ}$ & 17$5^{\circ}$\\
    i & $2^{\circ}$ & $2^{\circ}$\\
    $\omega$ & 2${\pi}$/$T_{obs}$ & 2$\pi$/$T_{obs}$\\
    $S_{b1}$(15GHz) & 2.7\,$10^{-4}$ & 2.7\,$10^{-4}$\\
    $S_{b1}$(8GHz) & 3.0\,$10^{-4}$ & 3.0\,$10^{-4}$\\
    $S_{b2}$(15GHz) & 5\,$10^{-2}$ & 8.3\,$10^{-4}$\\
    $S_{b2}$ (8GHz) & 7.5\,$10^{-2}$ & 1.24$10^{-3}$\\
    quiescent-level & 5\,Jy & 5\,Jy\\
    \hline
    \end{tabular}
    \end{table}
    \begin{table}
    \caption{Epochs of the 15GHz peaks obtained by the model-fittings.}
    \label{table:2}
    \centering
    \begin{tabular}{c c}
    \hline
    peak & epoch\\
   \hline
   secondary & 1962.35\\
    main &     1968.95\\
   secondary  & 1975.15\\
   main &    1981.75\\
   secondary & 1987.95\\
   main & 1994.55\\
   secondary & 2000.75\\
  main & 2007.35\\
   secondary & 2013.55\\
  main & 2020.15\\
   \hline
   \end{tabular}
  \end{table}
  \subsection{Results of model-fitting}
    Using model-1 we obtain the fitting for the 8GHz and
 15GHz lightcurves in
    Figures 2--6. Using model-2 we obtain the
 fitting for the 15GHz and 8GHz
    light curves in Figures 7-11.
    \begin{figure}
    \includegraphics[width=7cm,angle=-90]{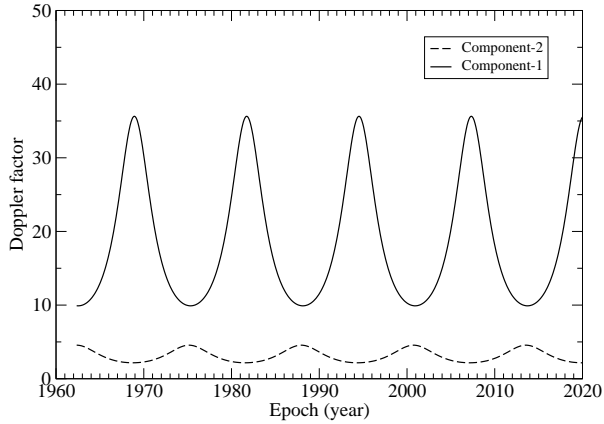}
    \caption{Model-A: The Doppler factors  of component-1 and
    component-2.}
    \end{figure}
    \begin{figure}
    \includegraphics[width=7cm,angle=-90]{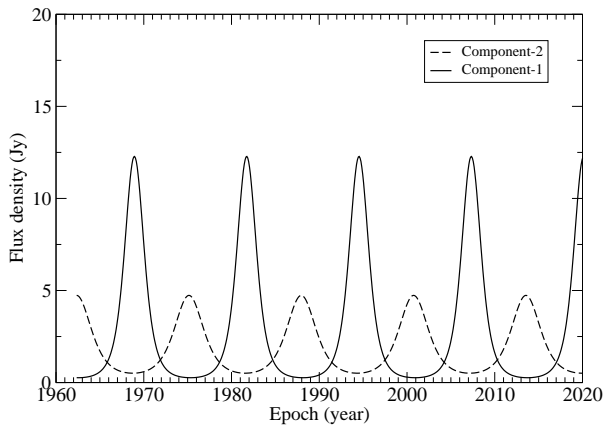}
    \caption{Model-A: the 15GHz flux light curves (Doppler profiles)
 of component-1 and -2.}
    \end{figure}
    \begin{figure}
    \includegraphics[width=7cm,angle=-90]{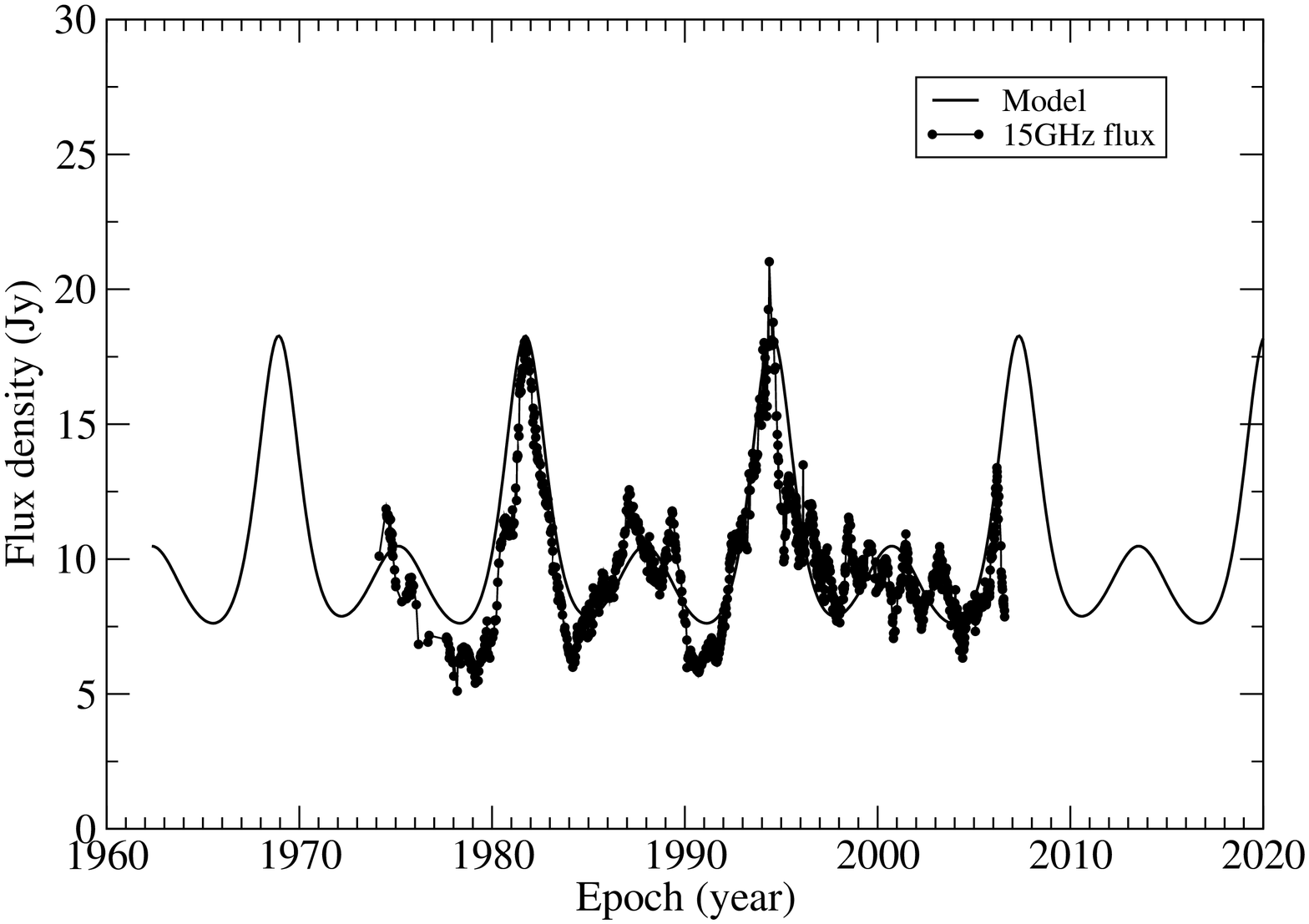}
    \caption{Model-A: model-fitting of the 15GHz lightcurve.}
    \end{figure}

     \begin{figure}
     \includegraphics[width=7cm,angle=-90]{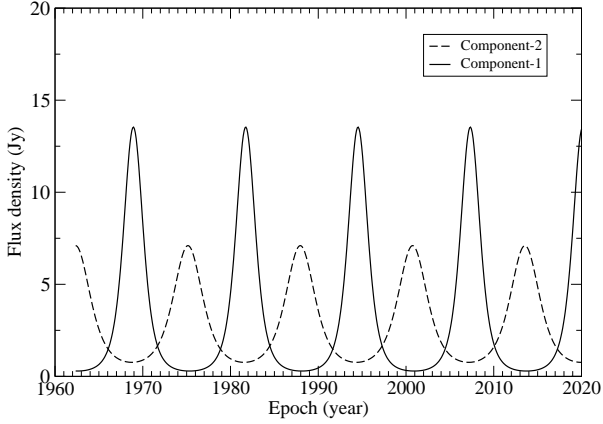}
    \caption{Model-A: the 8GHz flux light curves (Doppler profiles)
    of the component-1 and component-2.}
    \end{figure}
    \begin{figure}
     \includegraphics[width=7cm,angle=-90]{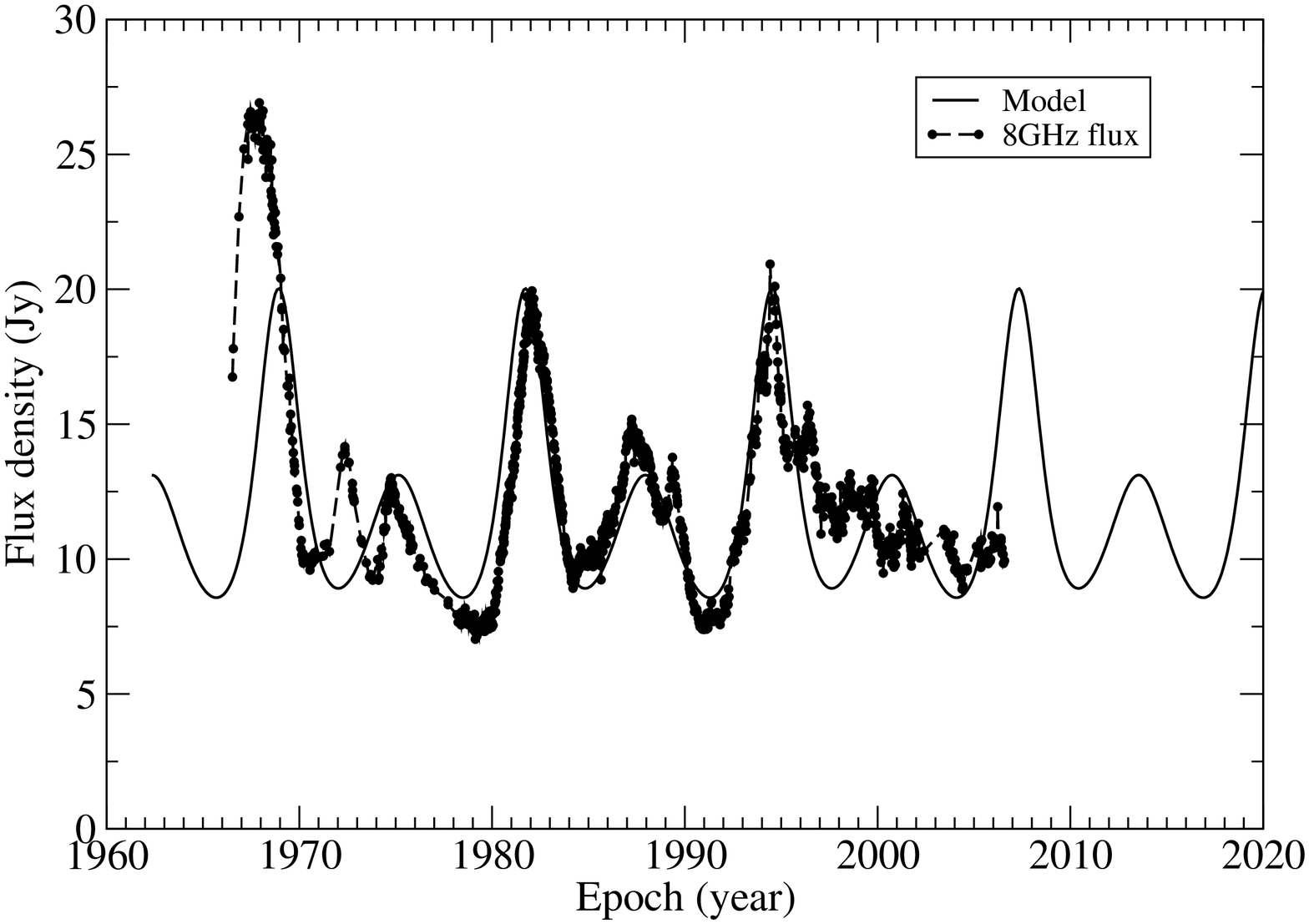}
     \caption{Model-A: model-fitting of the 8GHz light curve.}
    \end{figure}
    The peaks fitted and predicted for 15GHz lightcurve are shown in
    Table 2.\\
  Finally we summarize our results.\\
    \begin{figure}
    \includegraphics[width=7cm,angle=-90]{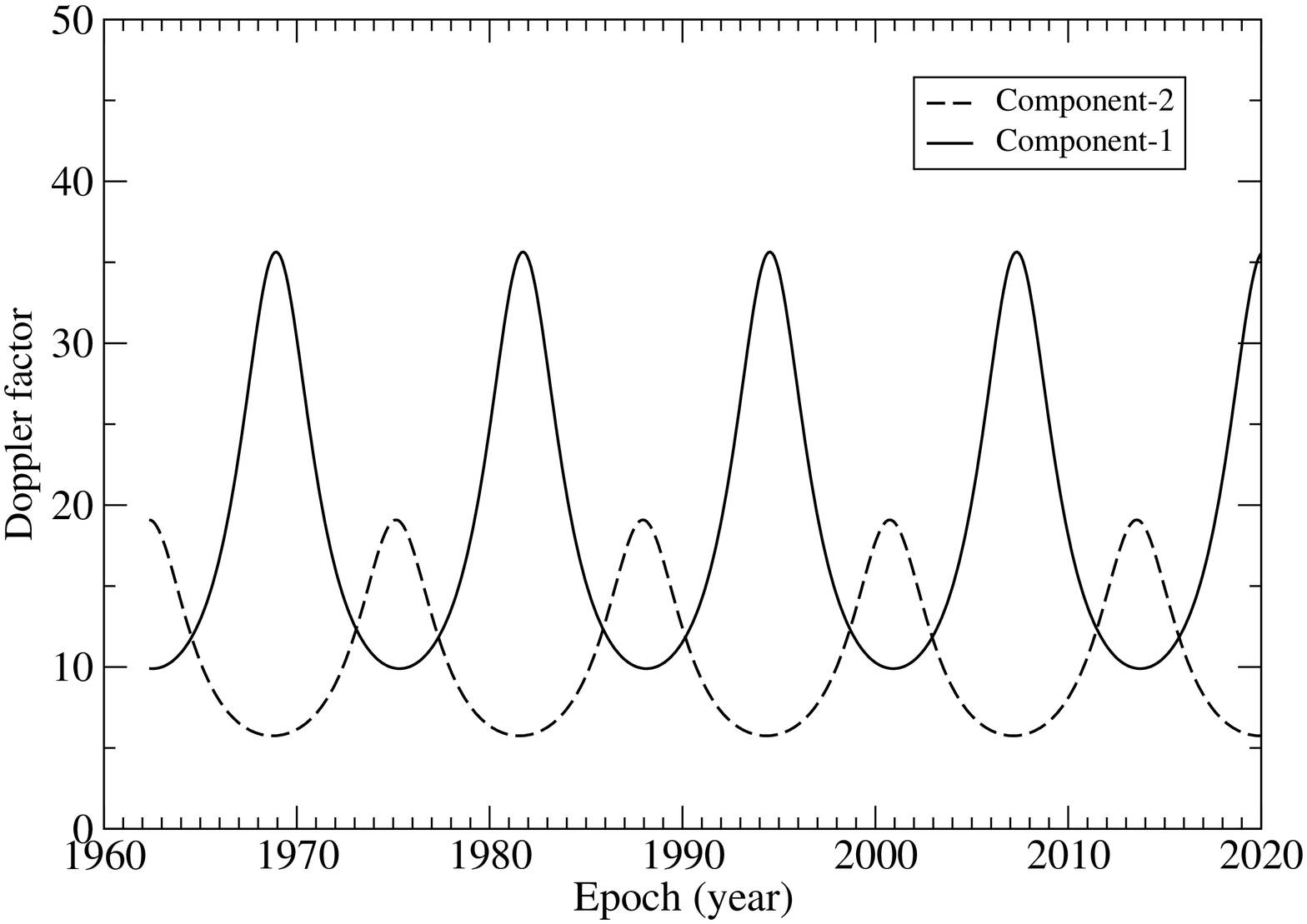}
   \caption{Model-B: Doppler factors of the component-1
   and component-2.}
   \end{figure}
  \begin{figure}
   \includegraphics[width=7cm,angle=-90]{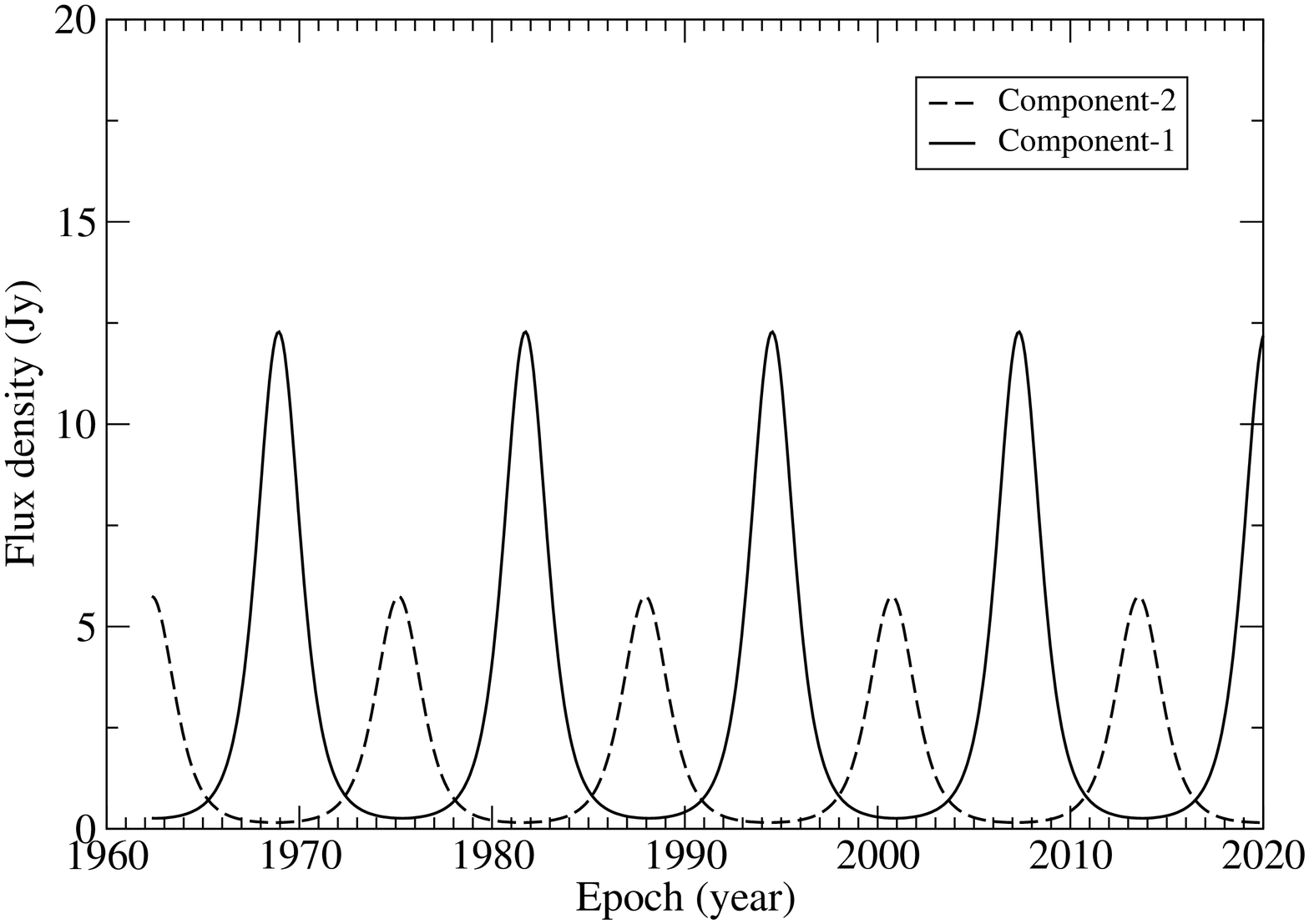}
   \caption{Model-B: 15GHz flux light curves (Doppler profiles)
   of component-1 and component-2.}
   \end{figure}
   \begin{figure}
   \includegraphics[width=7cm,angle=-90]{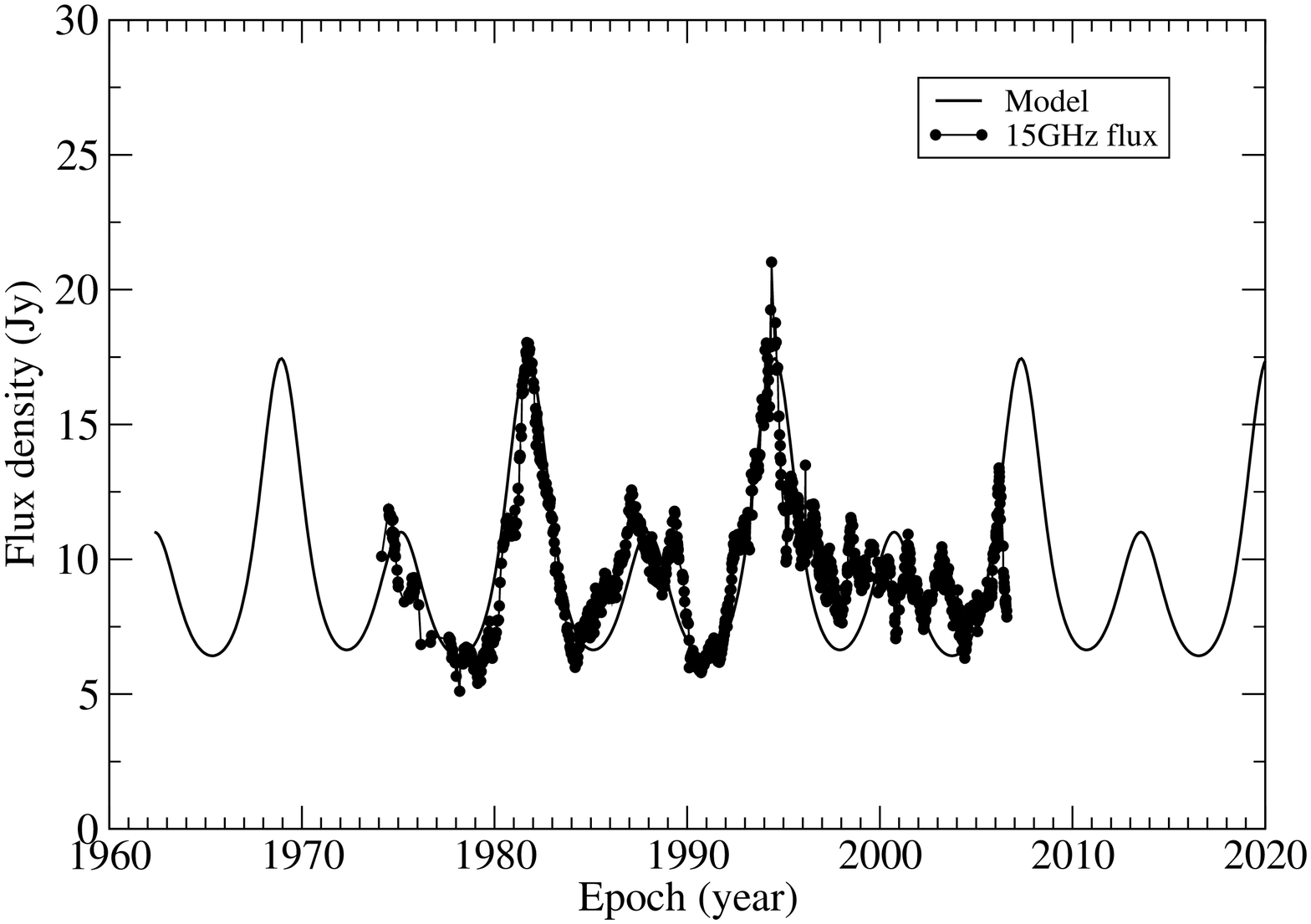}
   \caption{Model-B: model-fitting for the 15GHz light curve.}
   \end{figure}
   \begin{figure}
   \includegraphics[width=7cm,angle=-90]{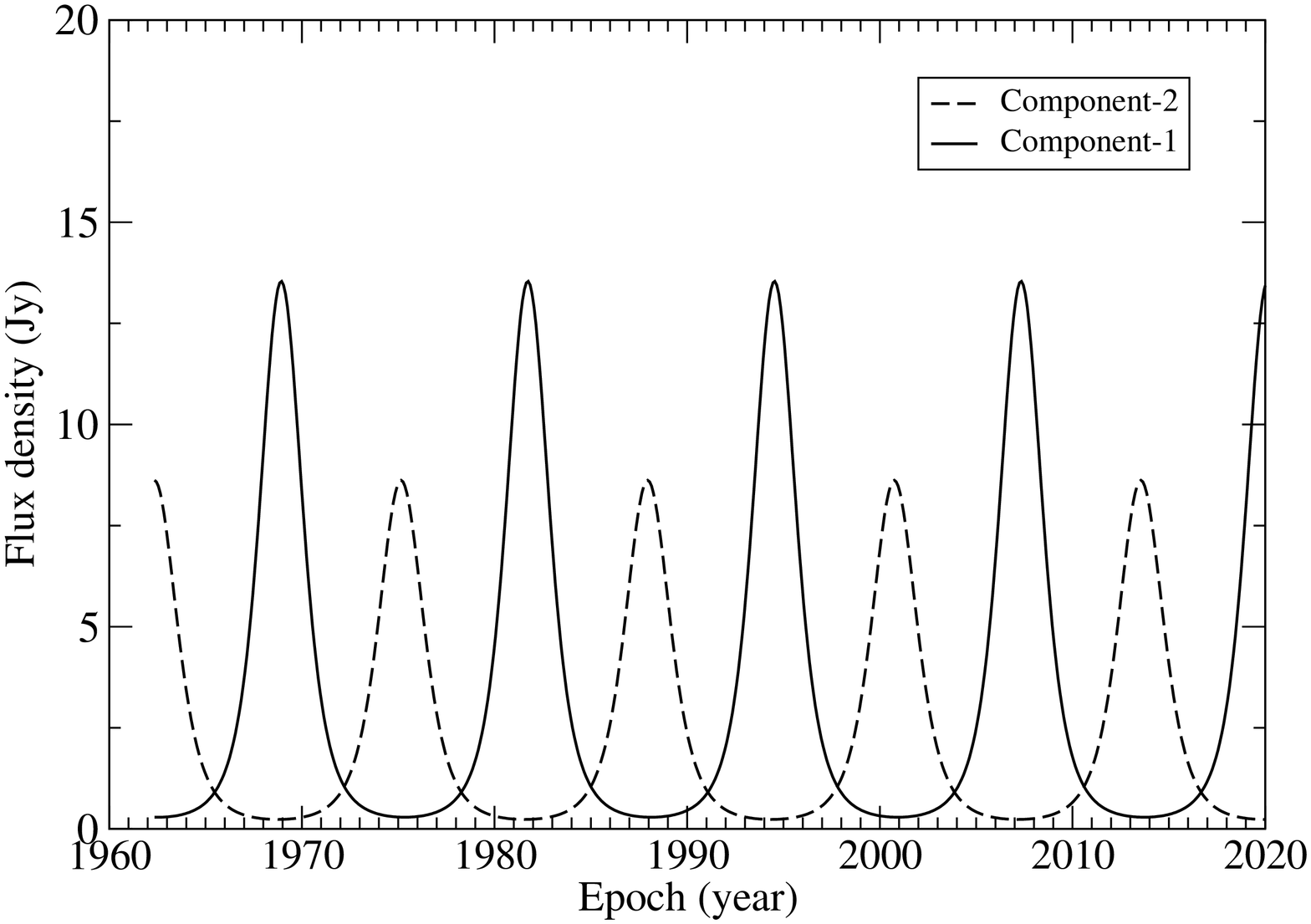}
   \caption{Model-B: 8GHz flux light curves (Doppler profiles)
  of component-1 and component-2.}
   \end{figure}
   \begin{figure}
   \includegraphics[width=7cm,angle=-90]{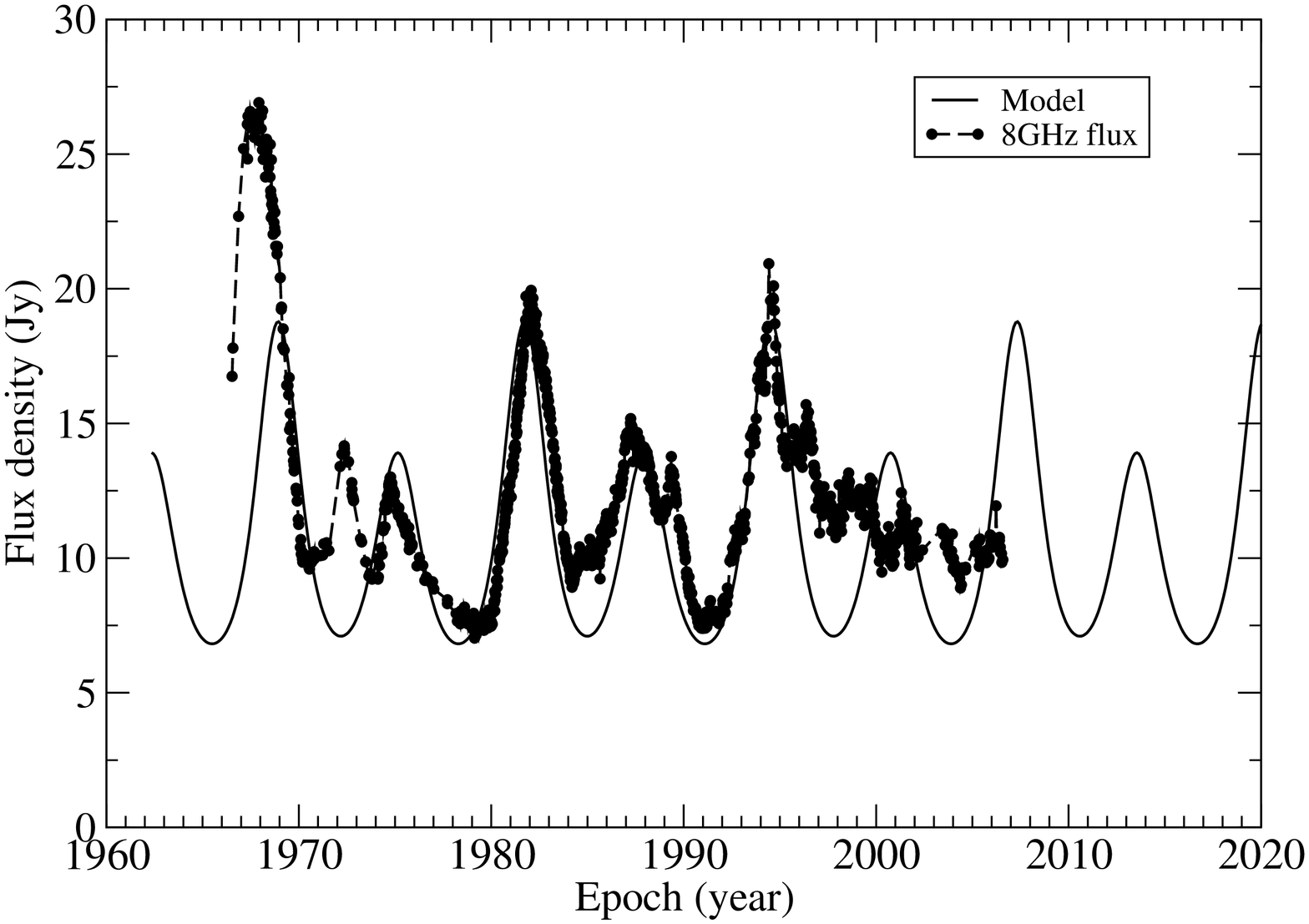}
   \caption{Model-B: model-fitting for the 8GHz light curve.}
   \end{figure}
  It can be seen that the model-fittings to the observed flux lightcurves
  are quite well. This implies that the periodic behaviour is dominated by
 the Doppler boosting effects (or the Doppler-boosting profiles) and the
  acrretion rates  and the mass-energy inflow into the jets both are
 inferred to be mostly
  stable during the period of $\sim$40 years. However,
   it can also be seen
  that deviations from the modelled lightcurves are apparent, e.g., the
 8GHz-peak in $\sim$1967.8 (Fig.4) and the behaviour
  observed at 15GHz in 2006 (Fig.6).
  We would attribute these
  deviations to be due to some wobble effects of
 the precessing jets or irregular variations in the accretion rates onto
  the black holes (and the energy-inflow into the jets).
 \subsection{Estimates for masses of black holes}
  In the paper we do not propose a specific binary black hole model to explain
 the dynamic properties of the jets, but we can roughly estimate
 the masses of the supermassive black holes. We assume that the precession of
  the jet axes is due to the orbital motion through a mechanism suggested
 by Roos et al. (\cite{Ro93}): when adding the orbital velocity of the hole to
 the jet velocity, the velocity vector of the jet precesses with the period
 equal to the binary period. Thus for a circular motion we have
 the formulae for the precessing angles of the primary and secondary
 black hole (cf. equations (3), (5)
 and (7) of Roos et al. (\cite{Ro93})):\\
 \begin{equation}
    {{\psi}_1}{\approx}{{2^{\circ}}.65}{\frac{m_8}{M_8}}
                                {(M_8+m_8)^{\frac{1}{3}}}
                     {(\frac{T_{int}}{yr})^{-\frac{1}{3}}}
                            {\cos}{{\chi}_1}/{{\beta}_{j1}}
 \end{equation}
 \begin{equation}
    {{\psi}_2}{\approx}{{2^{\circ}}.65}{(M_8+m_8)^{\frac{1}{3}}}
                     {(\frac{T_{int}}{yr})^{-\frac{1}{3}}}
                            {\cos}{{\chi}_2}/{{\beta}_{j2}}
 \end{equation}
 where ${\psi}_1$ and ${\psi}_2$ -- the angles of precession of the
 primary and secondary black hole (in units of degrees),
 $m_8$ and $M_8$ are the masses of the secondary and the primary
 black holes in units of $10^8$$M_{\odot}$, $T_{int}$ -- intrinsic
 orital period (= 6.9 yr.),
  ${\chi}_1$ and ${\chi}_2$ are the angles between the jet
 velocities and the orbital angular momentum (i.e. the normal of the orbital
 plane, in our case ${\chi}_1$=${\chi}_2$=0),
  ${\beta}_{j1}$ and ${\beta}_{j2}$-- the velocities of the jets in units
 of c (the speed of the light). If we take ${\psi}_1$=$5^{\circ}$
 and $m_8$$\approx$$M_8$ (${\beta}_j$$\sim$1), then we obtain $M_8$$\approx$
 22, i.e., $M_{BH}$$\approx$2.2\,$10^{9}$$M_{\odot}$. This estimated mass
 is consisitent with those obtained by other arguments (for example,
  an estimation (1.5\,$10^9$$M_{\odot}$)
  from the bolometric luminosity of 3C454.3 by
  Woo \& Urry \cite{Wo02}).

  \section{Conclusion}
 \begin{itemize}
  \item (1) the supposed  model is not unique. In addition to the
 double-jet models proposed in this paper, models with a single jet
  are also plausible. But in this case, the two emitting components
 could be assumed to be situated at different positio
ns in the
 curved jet and their rotation would produce the Doppler boosting
 profiles. In this paper we do not invoke a specific model for
 a binary black hole system.
  \item (2) In our model-fittings, multi-parameters can be chosen.
  When  parameters were chosen we consider mainly three ingredients: (a) the
  width of the two radio bumps; (2) the difference of Doppler factor at
  flaring and quiescent epochs should not be
 too large in order to avoid
 a huge difference in timescale of varibility at different epochs
 (Valtaoja et al. \cite{Va00});
  \item (3)  We should consider both effects due to Doppler
  boosting and the
   energy transfer into the jets (or accretion rates). In our models,
   the Doppler boosting determine the 'profiles' of the possible
  activity in the radio and optical bands.
  If the intrinsic (in the
  rest frame of the flows) strengths are stable, then the apparent activity
  follows the pattern of the Doppler boosting. This seems correct for
  the observed radio variations, for example during 1978--1995.
  \item (4) As for the optical variations, they do not follow the
 Doppler boosting profiles derived from our models. This could imply
 we should adopt different parameters for the optical variations, i.e.,
 the optically emitting regions could have orientations different
 from the radio regions. Moreover, optical flux variations could be
 largely determined by the variation in accretion rate and their
 much shorter timescales may be due to mechanisms for acceleartion of electrons
 and radiative losses.
  The optical peak at 2005.4 and radio peak at 2006.2, which both
 deviate the Doppler boosting pattern maximum, further indicate
 the necessity to consider the  effects both from
  Doppler boosting, accretion and transfer of enery into the jets.
 \item (5) The radio periodic behaviour still needs to be tested, because
  the 2005 optical flare has a quite peculiar mm-radio outbursts, which
 have much narrower profiles than before,
  very different from the 1981 and 1994 radio
  flare's profiles. This may imply that the radio
  periodic behaviour could be
 a short-term phenomenon. The optical-radio relation is still
 important to be tested and thus could obtain some information about
 the combination of the black hole disk system and the Doppler boosting effect.
 \item (6) Relationship between mm- and cm-outburtsts needs to be
 further studied
 in order to study the evolution of the outbursts.
 \item  (7) At present,  we cannot predict whether the periodic behaviour
 in 3C454.3 in radio and optical bands
  can hold long, and this should be observationally tested
  during the next period (2007--2020). Both optical and radio
  monitoring (intensity and polarization) are required.
 \end{itemize}
 \acknowledgements{S.J.Qian acknowledges the hospitality and financial support
 of the Max-Planck-Institut f\"ur Radioastronomie during his visit. N.A.
 Kudryavtseva was supported for this research through a stipend from the
 International Max-Planck Research School (IMPRS) for Rdaio and Infrared
 Astronomy at the Universitiies of Bonn and Cologne.}

 \end{document}